\documentclass[sigconf,10pt]{acmart}
\AtBeginDocument{%
  \providecommand\BibTeX{{%
    \normalfont B\kern-0.5em{\scshape i\kern-0.25em b}\kern-0.8em\TeX}}}

\copyrightyear{2022}
\acmYear{2022}
\setcopyright{acmcopyright}\acmConference[Morse'22]{1st ACM Workshop on AI
Empowered Mobile and Wireless Sensing }{October 21, 2022}{Sydney, NSW, Australia}
\acmBooktitle{1st ACM Workshop on AI Empowered Mobile and Wireless Sensing
(Morse'22), October 21, 2022, Sydney, NSW, Australia}
\acmPrice{15.00}
\acmDOI{10.1145/3556558.3558581}
\acmISBN{978-1-4503-9522-9/22/10}

\usepackage{pgfplots}
\usepackage{wrapfig}
\usepackage{graphicx}
\usepackage{subcaption}
\usepackage{array}
\usepackage{wasysym}

\usepackage{pgfplots}  
\pgfplotsset{width=6.6cm,compat=1.7}  

\usepackage{microtype}
\usepackage{amsmath}
\usepackage{multirow}

\usepackage{mathtools, nccmath}
\usepackage{lipsum}
\usepackage{graphicx}
\usepackage{caption}
\usepackage{subcaption}
\usepackage{ifthen}
\usepackage{comment}
\usepackage{multicol, blindtext}
\usepackage{ textcomp }
\usepackage{gensymb}
\usepackage{float}

\newcommand{\NAME}{\mbox{\textbf{L}i\textbf{T}alk}\xspace}

\begin{document}

\title{Detecting and Controlling Smart Lights with \NAME}
\author{Jagdeep Singh}
\authornote{Both authors contributed equally to this research.}
\affiliation{%
  \institution{Toshiba Europe Ltd.}
  \streetaddress{32 Queen Square}
  \city{Bristol}
  \country{UK}
  \postcode{BS1 4ND}
  }
\email{jagdeep.singh@toshiba-bril.com}
\author{Dan Watkinson}
\authornotemark[1]
\affiliation{%
  \institution{University of Bristol}
  \streetaddress{32 Queen Square}
  \city{Bristol}
  \country{UK}
  \postcode{BS1 4ND}
  }
\email{dan.watkinson@bristol.ac.uk}

\author{Tim Farnham}
\affiliation{%
  \institution{Toshiba Europe Ltd.}
  \streetaddress{32 Queen Square}
  \city{Bristol}
  \country{UK}
  \postcode{BS1 4ND}
  }
\email{tim.farnham@toshiba-bril.com}

\author{Daniele Puccinelli}
\affiliation{%
  \institution{University of Applied Sciences and Arts of Southern Switzerland (SUPSI)}
  \streetaddress{}
  \city{Lugano}
  \country{Switzerland}
  \postcode{}
  }
\email{daniele.puccinelli@supsi.ch}

\renewcommand{\shortauthors}{Jagdeep Singh \& Dan Watkinson}
\begin{abstract}
The rapid increase in demand for wireless controlled Smart Lighting has created a need to automate the mapping between the identifiers for individual light sources and their physical locations. To control Smart Lights, their IDs and physical locations relative to each other must be determined. Nowadays, skilled technicians perform this process manually, which requires a lot of effort, is time-consuming, and incurs high costs, particularly with non-stationary lights. Visible Light Communication has been presented as a possible solution to this problem. This paper presents an approach based on Visible Light Communication that leverages Machine Learning to automate the mapping process between the identifiers and the relative physical location of Smart Lights. We show that our approach provides a better location-mapping performance compared to existing methods.
\end{abstract}






\maketitle

\section{Introduction}
\label{sec:intro}
As light-emitting diodes (LEDs) become increasingly ubiquitous in commercial environments due to their energy efficiency, there has been a growing demand for wirelessly controlled Smart Lights. The Smart Lighting market was valued at \$11 billion in 2020 and is expected to reach \$18 billion worldwide by 2028~\cite{bloomberg_2022}. Aside from the clear advantages of remotely controllable lighting, Smart Lights can also enable location-based services through localisation techniques~\cite{luxapose}.
In order for LED-based Smart Lights to be integrated with control systems, their relative physical locations need to be known. In other words, a mapping between the identifier for each light and its actual location is needed. Currently, this process is carried out manually and is time-consuming, costly, and prone to human error. One solution to automate this process can be provided by Visible Light Communication (VLC), which uses the light in the visible spectrum to transmit information. With VLC, light can be used to encode and transmit the identification information to the camera system, which uses Machine Learning (ML) in order to map the identifiers to the actual light source location.
VLC-based localisation has benefits over radio-based approaches, including robustness to interference and multi-path fading~\cite{singh2020passive}.
VLC generally uses LED as transmitters and photodiodes or cameras as receivers. LED-to-Camera communication is a more recent form of VLC that uses a CMOS (Complementary Metal Oxide Semiconductor) camera as a receiver rather than a more traditional photodiode. CMOS cameras are commonly available on a range of devices, making it possible to use commercial off-the-shelf devices to build a VLC receiver that can collect data from one or several sources simultaneously. The rolling shutter effect of a traditional CMOS camera can also be exploited in order to decode data that is being transmitted by the light sources~\cite{luxapose,rainbowlight}.
\newline
Camera-based VLC can be used to build a system that allows the mapping of new lights to their physical location. Subsequently, this information can be used to determine relative locations by other VLC receivers; it should be noted that most of the work done in state-of-the-art VLC localization approaches is based on the assumption of such prior knowledge. This paper focuses on the initial determination of the relative locations of the lights to enable lighting control and support other location-based services.
\begin{figure}
 \vspace{-1.00mm}
    \centering
    \includegraphics[width=\columnwidth]{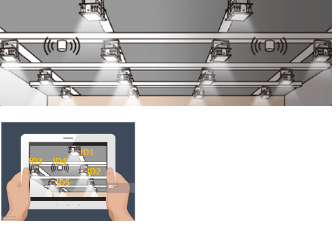}
    \caption{\NAME: VLC-based collection \& mapping of light IDs.}
    \label{fig:id_issue}
    \vspace{-5.00mm}
\end{figure}

\NAME is an initial step in this direction to explore the use of VLC to automate the process of simultaneous mapping and localisation for Smart Lights by predicting the link distance between light sources and camera receivers using Machine Learning. Based on the predicted link distance, the location coordinates of the light sources can be assigned, eliminating the assumption that light source positions are known a priori. With \NAME, we show that the use of Machine Learning can provide benefits compared to existing VLC decoding methods.
Figure~\ref{fig:id_issue} depicts the \NAME mechanism, where with the deployment of \NAME solution, the technician or autonomous guided vehicles (AGVs) can point a camera towards the installed Smart Lights, which illuminates/transmits their ID. The camera will decode those IDs and further assign the location coordinates to the lights based on the determined link distance.
In a nutshell, this paper makes the following contributions:
\begin{enumerate}
    \item A reliable many-to-one VLC link that can simultaneously decode the ID of multiple lights.
    \item A fast decoding pipeline for the above-mentioned VLC link, reducing the computation time to an average of 500 ms to determine the identifier of the lights compared to existing methods.
    \item A method to determine the link distance between the light and the camera based on the LED image captured by the camera. More specifically, we train a model to predict the link distance given the captured LED size within the camera frame.
\end{enumerate}
\section{\NAME Overview}
\label{sec:overview}
In this section, we will first present the modulation scheme we use in \NAME for the transmission of the light IDs and further discuss the image processing steps needed to decode those IDs from multiple lights simultaneously.  
\subsection{Transmitter}
In order to transmit data at a relatively high frequency, any transmitter must be capable of switching its light state at relatively high speeds. Most LEDs can support this fast switching as long as they are connected to a suitable driver circuit. VLC can also be carried out using an LED display as a transmitter\cite{screen-camera}. Data in VLC is most commonly transmitted by modulating the state of light to on or off depending on the binary data being transmitted. 

\subsubsection{\textbf{Modulation Scheme}} In \NAME, basic On-Off Keying (OOK) is used as the modulation scheme for the data. OOK works by representing a binary `0' by turning the light off and a binary `1' by turning it on. Using this method without any additional data manipulation will cause noticeable flickering if many consecutive `0' or `1' symbols were to appear in the data. In order to overcome this, we combine OOK with Manchester encoding of the data. Manchester encoding uses two bits to represent one symbol from the original data. In our case, a `0' is represented by the sequence `10' and a `1' is represented by the sequence `01'. This encoding strategy creates a balanced amount of ones and zeros in the transmitted data that keeps the light at a constant intensity. As it is not possible to have more than two consecutive symbols, this technique results in the absence of any noticeable flicker.

We use packets with a fixed preamble size and variable payload size. We use 5 symbols `10001' as the preamble (the use of 3 consecutive zeros makes it distinguishable from the rest of the data). A start frame delimiter of `01' is also used after the preamble before the payload.

The total size of the packet is calculated using the size of the preamble and the size of the payload. The size of a packet carrying $N$ bits of data is given by Equation \ref{eq:packet}.

\begin{equation}
    \mathrm{Packet~Size} = M + S + N * n_{\mathrm{symbol}}
    \label{eq:packet}
\end{equation}
where ``M'' represents the size of the preamble, ``S'' is the size of the start frame delimiter and ``N'' is the number of data bits in the payload. ``$n_{\mathrm{symbol}}$'' is the number of symbols that represent a single data bit after encoding (in our case $n_{\mathrm{symbol}} = 2 $).

\subsubsection{\textbf{Modulation Frequency \& Packet Size}}
For successful detection of information bits when using the camera as a receiver, the amount of information encoded in a packet also depends on the transmission frequency. We tested various transmitter frequencies in order to determine the most suitable frequency that maximises the amount of information encoded without creating too much decoding difficulty. Theoretically, there is a maximum transmitter frequency of that, which is half the camera scan rate. This theoretical upper
bound would cause a band to have a width of only one pixel on the image. This would make it difficult to decode.
In \NAME, the chosen frequency value is 2.5 kHz, this produces a band width, $W$ of 12 pixels. Modulating at this frequency causes no perceivable 
flicker and allows
for easy decoding of the data. It is calculated that the maximum expected band of 
the preamble will have a width of 36 pixels. Given this pixel width and our calculated packet size from Equation~\ref{eq:packet}, we can determine the total length in pixels of a packet in a frame.
\vspace{-6.00mm}
\begin{equation}
   \vspace{-1.00mm}
   \mathrm{Packet~Length~in~Pixels} = 12*23 = 276 \mathrm{~pixels}.
   \label{eq:packet_length}
   \vspace{-1.00mm}
\end{equation}

\subsection{Receiver}
\label{sec:cameraParams}
Any hardware that can detect the presence of light can be used as a receiver. Traditionally, photodiodes are preferred due to their fast response and high bandwidth. Some researchers have investigated the use of an LED in reverse bias mode \cite{ledtoled} and others have used smartphone cameras \cite{led-cmos}. Cameras have much better spatial resolution than photodiodes so can give advantages when decoding from multiple transmitters or carrying out localization\cite{luxapose}. In \NAME, we use the Raspberry Pi Camera to decode the ID of multiple lights simultaneously.

\noindent
\textbf{Rolling Shutter:}
Most traditional cameras contain either a CMOS or a \emph{charged-coupled device} (CCD). CCD-based cameras most commonly use the \emph{global shutter} method of capturing images from a scene. This is when all pixels are captured simultaneously so the frame taken represents a single instant in time. On the other hand, most CMOS-based cameras capture images with the \emph{rolling shutter} method. This is an alternative to a global shutter, where an image is taken by scanning across the scene either horizontally or vertically. A rolling shutter mechanism works by exposing each row or column of pixels sequentially. Once all the rows have been exposed, an image is created by merging them together. This means that not all parts of the image are recorded at the same instant in time. Predictable distortions can be produced when capturing fast-moving objects or rapid flashes of light.\\
When building a system that exploits the rolling shutter effect, it is important to carefully choose the camera parameters to obtain the correct \emph{exposure time} as this has a significant effect on the width of the light bands and, consequently, on the ability to decode.

\begin{figure}[h]
    \centering
       \vspace{-1.00mm}
    \begin{subfigure}[b]{0.4\columnwidth}
        \centering
        \includegraphics[width=\linewidth]{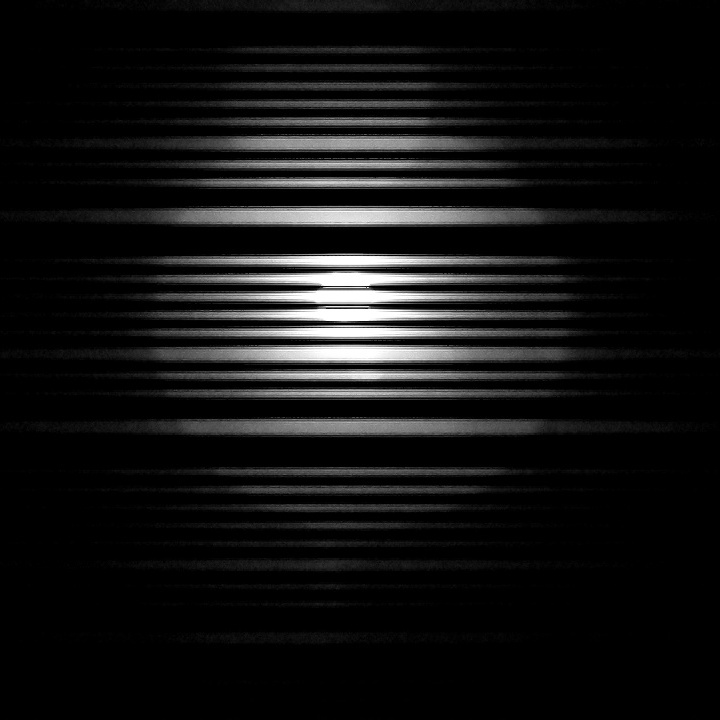}
        \caption{Original Frame \newline}
    \end{subfigure}
    \hfill
    \begin{subfigure}[b]{0.4\columnwidth}
        \centering
        \includegraphics[width=\linewidth]{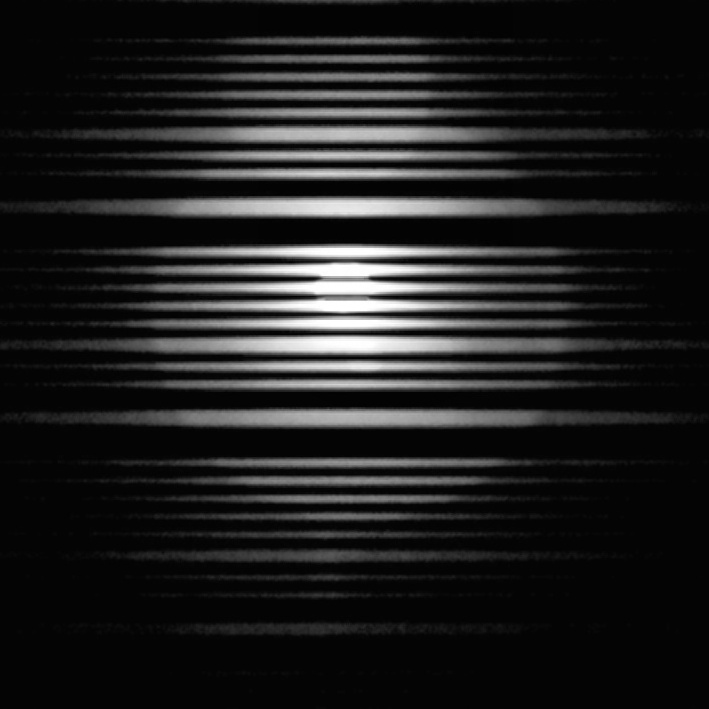}
        \caption{Contrast increase and 3x3 blur}
    \end{subfigure}
    \par\bigskip
    \begin{subfigure}[b]{0.4\columnwidth}
        \centering
        \includegraphics[width=\linewidth]{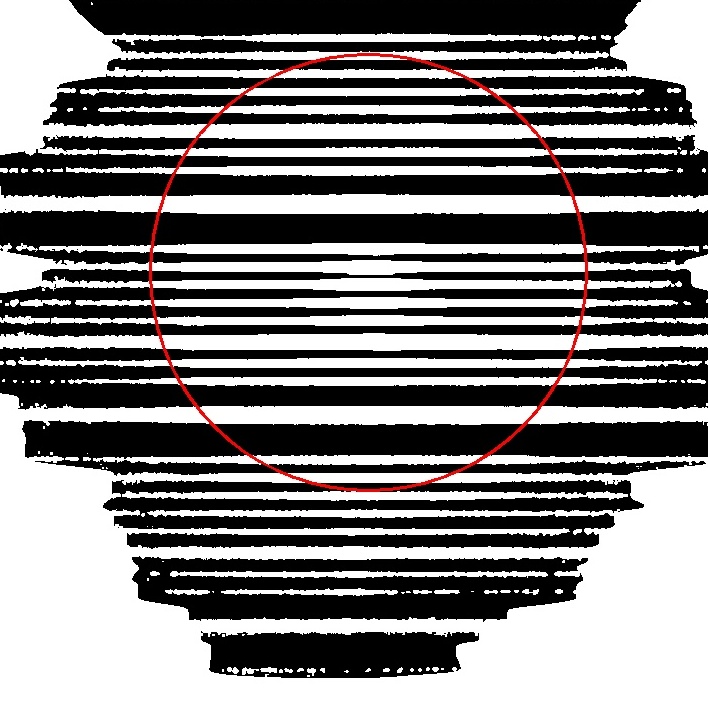}
        \caption{Adaptive threshold}
    \end{subfigure}
    \caption{\textbf{Decoding pipeline. } After the contrast increase, we blur the image (b) and apply an adaptive threshold to create a binary image (c); the red circle shows the detected radius}
    \label{fig:decoding}
    \vspace{-4.00mm}
\end{figure}

\subsubsection{\textbf{Camera Parameters}}

When designing a receiver for an LED-to-Camera system there is a range of aspects to consider. This section describes the most important parameters and how they should be set based on the design of the system.

\noindent
\textbf{Exposure Time:} It is defined as the time that the camera sensor is exposed to the light and is arguably the most important parameter. The exposure time closely relates to the rolling shutter speed of the camera, as previously mentioned. A shorter exposure time will create more definition between the dark and light bands, making them easier to distinguish. 
\citet{luxapose} show that independent of film speed (ISO), the best performance is achieved with the shortest exposure time. In \NAME, by using a camera connected to a single board computer rather than a smartphone, we have more control over the exposure time of the camera and can set a specific value to reduce the setup needed before transmission.

\noindent
\textbf{Film Speed:} It is a measure of the sensitivity of the image sensor to light. It can be represented as the amount of photons needed to saturate a pixel. An increased film speed (high ISO) will reduce the amount of photons needed to saturate a pixel. Usually this is used in low light situations. In our case, a slower film speed is preferred, because it increases the definition between bands and reduces noise.

\noindent
\textbf{Scan Rate and Resolution:} Due to the rolling shutter effect, the frames are captured row by row in a sequential way. The image resolution determines the number of rows in a frame. The scan rate is known as the total time it takes for the sensor to scan all the rows of the frame. The \emph{camera scan rate} can also be thought of the sampling rate of the sensor. The total scan rate of the camera is determined by the time it takes to scan a single row of pixels. A resolution of 1280x720 px was chosen for the images. It was decided not to use
Full HD (1920x1080) in order to reduce processing time and memory requirements.
The readout time $T_r$ is the time it takes for the sensor to read a single row of pixels. In \cite{rollinglight}, it is shown that the width $W$ of each band of light can be given by Equation \ref{eq:rolling}, where $f$ is the frequency of modulation.

\begin{equation}
    W = \frac{1}{2f T_r}
    \label{eq:rolling}
\end{equation}
\noindent
Equation \ref{eq:rolling} indicates that the width of the light bands does not depend on either the size of the LED or the distance from the transmitter. With the current parameters \& using Equation~\ref{eq:rolling}, the readout time calculated as 16.67$\mu$s.

\subsubsection{\textbf{Image Processing}}\label{image_process}
The receiver uses image processing techniques in order to detect the transmitter and decode the data.
We use the image processing pipeline as in \break Luxapose\cite{luxapose}. An additional contrast increase is used as in DynaLight \cite{dynalight}.

\noindent
\textbf{Decoding:} We adapt the method proposed in DynaLight\citet{dynalight} to decode the images captured by the camera. This includes the steps of blurring with a 3x3 kernel and then using an adaptive threshold to create a binary image.
In the center of the detected area, where the light is located, the intensity is much higher than in other parts. This causes some of the pixels in this location to become overexposed and create bands that are larger than those expected, making it harder to decode. In order to overcome this, we implement a row offset method. Instead of selecting the center column of pixels for decoding, a column offset of half of the circle radius is used for selection.

\section{Implementation}
\label{sec:implementation}

\begin{figure}[t!]
  \centering
  \includegraphics[width=0.6\columnwidth]{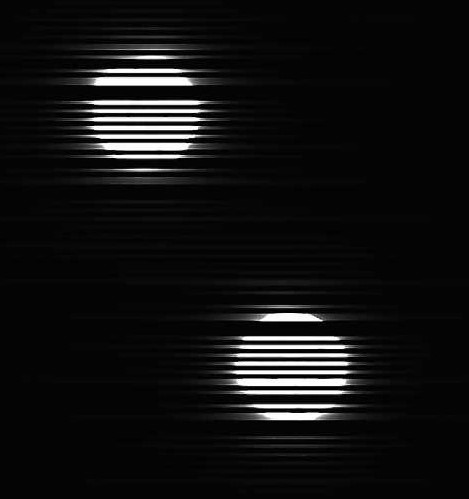}
   \vspace{-1.00mm}
  \caption{Captured LEDs using rolling-shutter in a single frame.}
  \label{fig:two_leds}
   \vspace{-4.00mm}
\end{figure}

To evaluate the performance of our \NAME design, we implement a prototype system using a transmitter circuit with off the shelf components and a commercially available single board computer together with a camera module. 

\noindent
\textbf{Transmitter:} On the transmitter side, we use an Arduino Uno R3 to perform modulation, control and encoding using an LED and a custom circuit. We can achieve the desired frequencies by using the built-in delay functions to create time periods between the on and off states of the light. We use this to achieve a modulation frequency of 2.5 kHz. The transmitter always repeats the same packet on a loop until deactivated. The only information transmitted is the hardware identifier, so a simple solution is to repeat packet transmission until the receiver can carry out decoding to avoid the need for synchronisation across the link.
The circuit used has been adapted from the circuit used in DynaLight \cite{dynalight}. We employ the X-Lamp MC-E by Cree, which contains 4 LED dies on an integrated circuit. The maximum current for each die is 700 mA.

\noindent
\textbf{Receiver:} Originally, the implementation was started using a specialised camera designed for machine vision applications, the OpenMV\footnote{https://openmv.io/products/openmv-cam-h7-plus}. After some testing with the camera, we realized that the built-in processor was not powerful enough for the required image processing techniques.
The final solution employs a Raspberry Pi 4 Model B single board computer. This is connected to the Raspberry Pi HQ camera\footnote{https://www.raspberrypi.com/products/raspberry-pi-high-quality-camera/}. The Raspberry Pi contains a Broadcom Quad-core 1.5 GHz processor, making it much more powerful than the OpenMV's single core 480 MHz processor. Using the Raspberry Pi allows us to capture and process images on the same device, removing the undesired USB serial latency that was occurring. The Raspberry Pi runs its own operating system based on Debian, which provides all the necessary software that is utilised in the system.

\subsection{Mapping Method} \label{sec:map_method} The first step for mapping the light ID to their physical location coordinates, with respect to the camera, is to know the elevation angle or the link distance between the camera and light source. In \NAME, we use the following approach to calculate the link distance.\\  
\textbf{Method 1 - Conventional decoding:} In rolling-shutter-based LED-to-Camera communication, the number of bright pixels that appear on the image depends on the link distance between the light source and camera. Mathematically taken from~\cite{dalgic2022enabling},  the relation between distance, number of pixels and camera factors can be represented as:
\begin{equation}\label{eq:dist_conv}
    D_{\mathrm{link}}=f_L+\frac{S_r f_L}{S_{ip}S_p}
\end{equation}
where $D_{\mathrm{link}}$ is the link distance between the camera and LED, $f_{L}$ is the focal length of the camera (equals to 0.28 mm in our case). $S_{r}$ is the radius of the LED (taken as 10.9 mm in \NAME), $S_{ip}$ is the size of LED in terms of pixels and $S_p$ is the length of a single pixel. In \NAME, we use Equation~\ref{eq:dist_conv} to determine the link distance from two light sources simultaneously.\\
\textbf{Method 2 - \NAME:} We employ a linear regression model to calculate the $D_{\mathrm{link}}$ to improve the link distance estimation accuracy. We train the model for different sizes of LED received on the image sensor and for different known link distances. The training was performed in a dark room with no ambient light sources.
During the evaluation, we first process the image of the received light (an example using two LEDs is shown in Figure~\ref{fig:two_leds}), calculate the blob size, i.e. LED size appears on the image, and predict the distance.
 
Based on the calculated distance from two LEDs and with the known camera location and separation between the two LEDs, the location coordinates to LEDs can be calculated and assigned to map the lights to their physical location. 
In \NAME, we have only implemented the former part, and the latter part is left for future work. 
The results and evaluation for distance estimation using two methods are described in Section~\ref{sec:exp}.

\section{Evaluation}
\label{sec:exp}

\begin{figure}
    \centering
    \vspace{-1.00mm}
    \includegraphics[width=0.8\columnwidth]{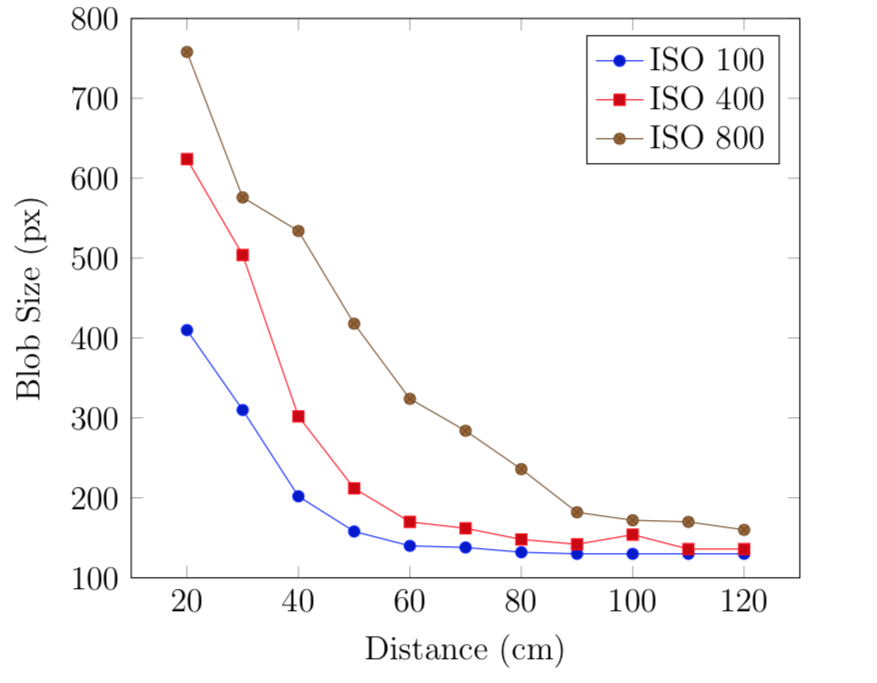}
    \caption{\textbf{Blob Size vs Distance and ISO.} Each line represents a different ISO (film speed) value.}
    \label{fig:blobsize}
    \vspace{-4.00mm}
\end{figure}

\begin{figure*}[h!]
	\centering
	\begin{subfigure}[t]{0.33\textwidth}
    	\centering
    	\includegraphics[width=\columnwidth,height=4cm]{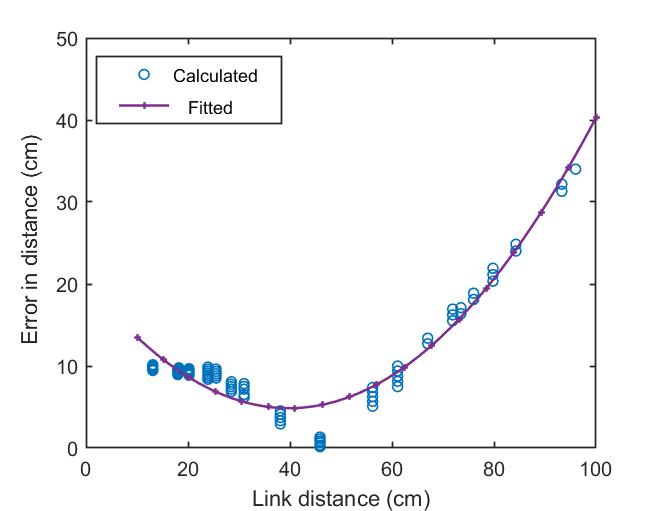}
    	\caption{Error in link distance using conventional mathematical method.}
    	\label{fig:conv_method}
    \end{subfigure}	
    \begin{subfigure}[t]{0.33\textwidth}
    	\centering
    	\includegraphics[width=\columnwidth]{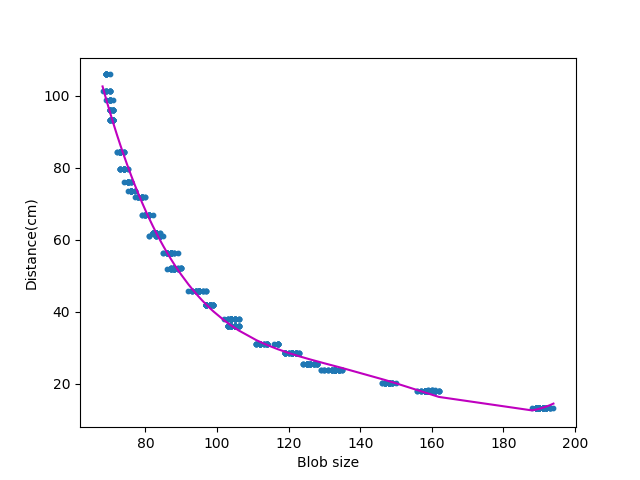}
    	\caption{ML predicted distances for given LED size.}
    	\label{fig:ML_predict}
    \end{subfigure}    
    \begin{subfigure}[t]{0.33\textwidth}
    	\centering
    	\includegraphics[width=\columnwidth]{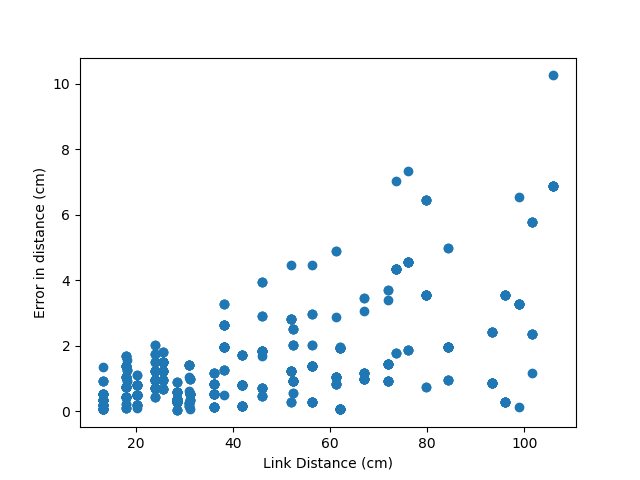}
    	\caption{Measured error in link distance using regression.}
    	\label{fig:err_ML}
    \end{subfigure}
    \caption{\NAME link distance analysis.}
    \label{fig:dist_results}
     \vspace{-3.00mm}
\end{figure*}

In this section, after a discussion of how film speed can affect transmissions over distance and how it can be best tuned, we evaluate the distance estimation of the prototype system. 

\subsection{Transmitter}
\textbf{Duty Cycle:} Our preliminary experiments showed that the dark bands on the frame produced by the camera are much thinner than the bright bands due to the extra time needed to charge and discharge the LED. We addressed this issue by making the off time slightly larger than the on time, i.e., by changing the duty cycle of the signal to 40\%.
\subsection{Receiver}
\subsubsection{\textbf{Effect of Film Speed on Distance}}\label{sec:filmSpeed}
Our preliminary experiments with the complete implementation showed that the system found it more difficult to decode data at longer distances when using a fixed length packet. This is because, as the distance between the transmitter and receiver increases, the blob size provided by the detection pipeline decreases. Therefore, there are fewer pixels to decode in an image, and using a fixed length packet imposes a lower bound on the blob size to enable decoding. 
The authors of Luxapose \cite{luxapose} state that it is best to keep film speed to a minimum to ensure reliable decoding. They mention that increasing the film speed can help the decoding process at longer distances but is conducive to images with more noise. Therefore, it is important to look into the relationship between film speed (ISO), distance and the blob size. 

To this end, we carry out an experiment using three different ISO values; 100, 400 and 800. The distance between the receiver and the transmitter is measured, and the diameter of the blob is recorded. 
Figure \ref{fig:blobsize} shows that, as the distance is increased, the blob size decreases, making size estimation less accurate. 
This decrease in blob size can be partially overcome by increasing the film speed of the camera. It is seen that using an ISO value of 800 results in the maximum blob size at all distances.
We can also calculate a lower bound of the blob size due to the fixed packet size i.e., 276 pixels (using Equation~\ref{eq:packet_length}). The graph shows that when using an ISO value of 100 or 400, the blob size becomes less than that value at around 40 cm, whereas when using an ISO value of 800, this point is not reached until around 70cm. As the performance of \NAME for distance estimation depends on the blob size, optimising the camera parameters such as ISO evaluated in this section is vital.

\subsection{Performance Analysis}
\subsubsection{\textbf{Computation Time}}
Our experiments show that \NAME takes between 0.3-1.3 seconds to decode a frame (containing two LEDs) with an average time of 500 ms. This is much lower than the total time taken by other systems such as Luxapose~\cite{luxapose}. The time to decode varies based on the scheduling methods of the kernel running on Raspberry Pi.  

\subsubsection{\textbf{Distance estimation}}
Using conventional decoding methods (Method 1 as described in Section~\ref{sec:map_method})~\cite{dalgic2022enabling}, our analysis shows that the error is minimized only at a specific distance, i.e. when the received LED image has no blooming effect and no blurriness. Further, the error is inconsistent and increases with distance. For shorter link distances, the average error is more than 2 cm and the pattern in the error plot for distance estimation is not linear, as illustrated in Figure~\ref{fig:conv_method}. The mean squared error calculated with this approach is greater than 10cm.

In Section~\ref{sec:filmSpeed}, we have seen how the camera parameters and link distance affect the blob's size. This implies that it is possible to estimate the link distance based on the blob size. We employ our regression model (Method 2 as described in Section~\ref{sec:map_method}) to perform the analysis at various distances and predict the link distance based on the received blob size while simultaneously decoding the light ID. The pink curve in Figure~\ref{fig:ML_predict} shows the predicted distance, and the error in the predicted distance for various positions is shown in Figure~\ref{fig:err_ML}. The mean squared error calculated using ML model is 1.9734 cm, resulting in a five-fold improvement compared to the existing method (see Figure~\ref{fig:conv_method} \& \ref{fig:err_ML}). Further, performance comparison with the SOA work can only be made after full implementation of \NAME for localization service as the estimation distance is the initial step. As part of our future work, we plan to train the model to decode the LED IDs, identify the link distance within a frame, and provide location services.
\section{Conclusion}
\label{sec:con}
With an increased demand for wireless LED control, camera-based VLC proves to be a potential solution to the ID matching problem. This is defined as the requirement to match hardware identifiers of light fittings to their locations. In this paper, we have proposed a way to address this problem using off-the-shelf available cameras and a link distance estimation regression model. Our results show a five-fold improvement in the distance estimation mean squared error compared to existing methods. Furthermore, we have developed a fast decoding pipeline to reduce the frame decoding time to an average of 500 ms. 

 \section*{Acknowledgement}
 This work has been funded by the European Union's Horizon 2020 research and innovation programme under the Marie Skłodowska Curie grant agreement ENLIGHT'EM No 814215.
 
\bibliographystyle{ACM-Reference-Format}
\bibliography{references}

\end{document}